\documentclass[runningheads]{llncs}
\titlerunning{Program Semantics and a Verification Technique for Knowledge-Based Multi-Agent Systems }
\title{Program Semantics and a Verification Technique for Knowledge-Based Multi-Agent Systems }
\author{
	F.~Belardinelli\inst{1} \and
	I.~Boureanu\inst{2} \and
	V.~Malvone\inst{3} \and
	S.~F.~Rajaona \inst{2}}

\institute{
	Imperial College London\ \email{francesco.belardinelli@imperial.ac.uk} \and
	University of Surrey\ \email{\{s.rajaona,i.boureanu\}@surrey.ac.uk}\and
	Télécom Paris\ \email{vadim.malvone@telecom-paris.fr} 
}
 \usepackage{graphicx}
\usepackage[hyphens]{url}
\usepackage{hyperref} 
\usepackage[normalem]{ulem}
\usepackage{amsmath}
\usepackage{paralist}
\usepackage{amsmath}
\usepackage{booktabs}
\usepackage{amsfonts}
\usepackage{amssymb}
\usepackage{xcolor} 
\usepackage{mathtools}

\renewcommand{\wp}{\mathit{wp}}
\newcommand{\SP}{\mathit{SP}}

\newcommand{\agset}[1]{#1}

\def\qed{\hfill{\qedboxempty}      \ifdim\lastskip<\medskipamount \removelastskip\penalty55\medskip\fi}

\def\qedboxempty{\vbox{\hrule\hbox{\vrule\kern3pt
			\vbox{\kern3pt\kern3pt}\kern3pt\vrule}\hrule}}

\usepackage{ifthen}

\renewcommand{\qed}{\hfill $\blacksquare$}  
\newcommand{\reason}[1]{\tag*{\small #1} }

\newcommand{\ioana}[1]{\color{blue} {IB: #1} \color{black}}
\newcommand{\fortunat}[1]{\color{black!36!green} {FS: #1} \color{black}}
\newcommand{\fb}[1]{\color{orange} {FB: #1 } \color{black}}
\newcommand{\vadim}[1]{\color{brown} {VM: #1 } \color{black}}

\renewcommand{\ioana}[1]{~}
\renewcommand{\fortunat}[1]{~}
\renewcommand{\fb}[1] {~}
\renewcommand{\vadim}[1]{~}

\newcommand{\nonobs}{\mathbf{n}}  
\newcommand{\obs}{\mathbf{o}}  
\newcommand{\new}{\mathbf{new}\ }  
\newcommand{\domain}{\mathsf{D}}
\newcommand{\dom}{\mathsf{dom}}
  
\newcommand{\eqdef}{\stackrel{\text{def}}{=}}
  
\newcommand{\If}{\mathbf{if}\ }  
\newcommand{\Then}{\ \mathbf{then}\ }  
\newcommand{\Else}{\ \mathbf{else}\ }  
\newcommand{\etal}{\textit{et al}. }  
  
\newcommand{\interp}[1]{[\![ #1 ]\!]}

\newcommand{\Skip}{\mathbf{skip}} 
 
\newcommand{\kv}{\mathrm{Kv}}

\newcommand{\modelsfo}{\models_{{}_{\mathit{FO}}}}  
\newcommand{\modelsqf}{\models_{{}_{\mathit{QF}}}}

\newcommand{\nexpr}{e}

\newcommand{\bk}{\backslash}

\newcommand{\pvar}{\mathbf{p}}

\newcommand{\var}{\mathcal{V}}

\newcommand{\nondet}{\sqcup}

\newcommand{\epistlang}{\mathcal{L}^m_{\mathit{K}}}
\newcommand{\folang}{\mathcal{L}_{FO} }  
\newcommand{\lqf}{\mathcal{L}_{QF} }  
\newcommand{\ldk}{\mathcal{L}^m_{\mathit{DK}}}  
\newcommand{\proglang}{\mathcal{C}}

 \newboolean{long}
\setboolean{long}{true}

\begin{document}

\maketitle

\begin{abstract}~
We give a relational and a weakest precondition semantics with perfect recall, for
concrete ``knowledge-based programs'', i.e., programs written in a general purpose programming
language but with tests on knowledge.
Using these knowledge-based programs, we define a program-epistemic logic
to model complex epistemic properties of the execution of multi-agent
systems.  
We give our logic a Kripke possible worlds semantics based on the
observability of variables so as to richly express
knowledge of agents who can or cannot observe said variables.  
We give a sound translation of the validity of program-epistemic logic
formulas into first-order validity, using our weakest precondition
semantics and an ingenious book-keeping of variable assignment.
We implement our translation 
and fully automate our verification method for well-established examples using SMT solvers.
\end{abstract}

\vspace*{-0.4cm}
\section{Introduction \& Preliminaries} 
\label{sec:intro}

Epistemic logics~\cite{hintikka62,fagin95a} formally look at the
dynamics of agents' knowledge, in properties such as information flow
in security protocols~\cite{BoureanuCohenLomuscio09} as well as
convergence in intelligent swarms of
robots~\cite{BoureanuKouvarosLomuscio16}.  In this work, we offer
a new method to verify such knowledge changes in the context of
programs, specifically in the interesting and realistic case when agents
observe only certain variables of the program space.
We first introduce a series of logic-related notions that are key in
positioning our work in the field, set out the scene, and explaining
our contributions.

\paragraph{\bf{Epistemic Logics.}} 
Logics for knowledge or epistemic logics~\cite{hintikka62} were first
 given a so-called Kripke, possible-worlds semantics. Assuming a set
 of agents, a Kripke structure (for epistemic logics) consists of a
 set of possible worlds, linked by
 an indistinguishability relation for each agent. Then,
 an \emph{epistemic formula $K_a \phi$}, which reads as ``agent $a$
 knows that $\phi$'', holds true at a world $w$, if the statement
 $\phi$ is true in all worlds that agent $i$ considers as
 indistinguishable from world $w$.

\textit{{Modelling Imperfect Information in Epistemic Logics.}} 
A possible-worlds semantics do not suffice to capture faithfully
aspects of distributed systems, in which -- for instance -- an agent has
private information that another does not.  To this
end, \emph{interpreted systems} were introduced~\cite{parikh85},
whereby agents are associated with private local states, which, put
together, make up the global state or world. Then, knowledge evaluated
over {interpreted systems} hinges closely on these local states: an
agent cannot distinguish two global states if her local states within
these global states are  equal.  
Several lines also looked
at how epistemic logic with
imperfect information could be expressed via notions of visibility (or
observability) of propositional variables, e.g.,~\cite{wooldridge01b,semi-public,charrier2016building}.

\textit{Logics of Visibility for Programs.} Others~\cite{IJCAI17,morgan06,rajaona} looked at how multi-agent epistemic logics with imperfect information would apply not to generic systems, but specifically to programs (over an arbitrary first-order domain), and -- with that -- at agents' visibility of  programs variables;  this transforms program's state space into a possible-worlds model.  In this domain, 
 the epistemic predicate $K_a(y=0)$ denotes that agent $a$ knows that
 the variable $y$ is equal to 0 (in some program). So, such a logic
 allows for the expression of knowledge properties of program states,
 using \emph{epistemic predicates}. This is akin to how, in classical
 program verification, one encodes properties of states using
 first-order predicates: e.g.,
 Dijkstra's \emph{weakest precondition
\footnote{The weakest precondition $wp(P,\phi)$ is a predicate such that: for any precondition
$\psi$ from which the program $P$ terminates and establishes $\phi$, $P$ implies $wp(P,\phi)$.} 
}
 ~\cite{dijkstra76}.

\textit{Perfect vs Imperfect recall.} For any of the cases aforesaid, an aspect often considered is the amount of knowledge that agents retain,
i.e., agents forget all that occur before their current state -- \emph{memoryless (or imperfect recall) semantics}, or agents  recall all their  history of states -- \emph{memoryful (or perfect recall) semantics}, or in between the two cases -- bounded recall semantics.

\paragraph{\bf{``Program-epistemic'' Logics}.}  To reason about knowledge change, epistemic logic
is usually enriched with \emph{dynamic modalities} from Dynamic Logics~\cite{Pratt+76a,Harel1984}.
Therein, a dynamic formula $\square_P\phi$
expresses the fact that when the program
$P$'s execution terminates, the system reaches a state satisfying
$\phi$ -- a statement given in the base logic (propositional/predicate
logic); the program $P$ is built from abstract/concrete actions (e.g.,
assignments), sequential composition, non-deterministic composition,
iteration and test, possibly.

Gorogiannis \etal~\cite{IJCAI17} gave a \emph{``program-epistemic'' logic},
which is a dynamic logic with concrete programs (e.g.,
programs with assignments on variables over first-order domains such as
integer, reals, or strings) and having an epistemic predicate logic as its base logic. 
Moreover, \`a la the
aforesaid~\cite{wooldridge01b,morgan06,rajaona}, the epistemic
model in~\cite{IJCAI17} relies on the observability of the programs'
variables by agents.  Gorogiannis \etal transformed program-epistemic
validity into a first-order validity problem, and in practice,
outperformed the then state-of-the-art tools in epistemic properties
verification.

\noindent
\paragraph{\bf{Our Work: Enhanced ``Program-epistemic'' Logics \& Its Verification}.}
We build on top of \cite{IJCAI17} in several significant ways. 
Firstly, the verification mechanisation in~\cite{IJCAI17} only
supports ``classical'' programs; this means
that~\cite{IJCAI17} cannot support tests on knowledge of
agents. Yet, such tests are essential in modelling communication
of knowledge: e.g., in epistemic puzzles such as the Muddy Children
and Cheryl's birthday~\cite{Lehman+84a} and in the so-called
``knowledge-based'' programs used to abstract communication
protocols~\cite{fagin95b}.
Secondly, the logic in~\cite{IJCAI17} allows only for knowledge reasoning after a
program $P$ executed, not before its run (e.g., not $K_{alice}
(\square_P \phi)$, only $\square_P (K_{alice} \phi)$). 
Lastly, the framework in~\cite{IJCAI17} assumes a single-agent epistemic logic
and does not allow for reasoning about nested knowledges operators
(e.g., $K_{alice} (K_{bob} \phi)$).

We lift all the limitations of~\cite{IJCAI17} listed above, by creating a new
``program-epistemic'' logic, with perfect recall and an observability-based semantics.
Our logic can be verified fully for formulas where the programs modality contains 
tests on knowledge, and for formulas with nested knowledge operators in the
multi-agent setting. This is naturally much more expressive, enabling the
 encoding and verification of a whole new range of problems in
computer science in our formalism.

The way perfect recall and imperfect information come together in our results is noteworthy.  On the one hand, is well-known that knowledge
under perfect recall and imperfect information is harder to
analyse~\cite{mcmas,atl-ferucio}. On the other hand, in our method, it is -- to a
great extent -- perfect recall that allows us to give a translation
from the validity of a ``program-epistemic'' logic into first-order
validity, by aiding us to tame assignments of variables under the
imperfect information (observability) of agents. In this sense, our
work can be seen as leveraging several elements of knowledge
semantics to obtain a new and surprisingly expressive verification
method for an epistemic logic on programs.

\paragraph{\bf{Contributions}.} 
We bring the following contributions:
\begin{enumerate}

\item We define a \emph{multi-agent program-epistemic logic} \emph{$\ldk$}, which is a dynamic logic whose base logic is a multi-agent first-order epistemic logic, under an observability-based semantics (Section~\ref{sec:logic}).

\item We give a knowledge-based programming language $\mathcal{C}$ (programs with tests on knowledge)
  that concretley defines the dynamic operators in $\ldk$. 
  We associate the programming language $\mathcal{C} $ with a  relational semantics and a weakest-precondition semantics, and we show their equivalence (Section~\ref{sec:programs}).

\item 
We give a sound translation  of the validity of a program-epistemic logic  
  into first-order validity (Section~\ref{sec:translation}).

\item We implement the aforesaid translation to allow a fully-automated verification 
  with our program-epistemic logic, via SMT-solving (Section~\ref{sec:implementation}).

\item  
  We verify the  well-known  Dining Cryptographer's protocol~\cite{Chaum1988} and the epistemic puzzle called the ``Cheryl's birthday problem''~\cite{van2017cheryl}. We report  competitive verification results. 
Collaterally, we are also the first to give SMT-based verification of the ``Cheryl's birthday problem''~\cite{van2017cheryl} (Section~\ref{sec:implementation}).

\end{enumerate}

\vspace*{-0.4cm}
\section{Logical Languages $\folang$ and $\ldk$} 
\label{sec:logic}

{We introduce the logics \emph{$\folang$}, \emph{$\folang$}, and \emph{$\ldk$}, 
used to describe states and epistemic properties of states, 
and program-epistemic properties of states.}

\subsection{Syntax of {$\folang$}, \emph{$\epistlang$}, and {$\ldk$}}

\paragraph{\bf{Agents and variables.}}
\label{sub:agents_and_variables}

We use \emph{$a,b,c, ...$} to denote agents,  \emph{$Ag$} to denote their whole set, and \emph{$\agset{G}$} for a subset therein. 

We  consider set  $\var$ of variables
such that each variable $x$ in $\var$ is ``typed'' with the group of agents that
can observe it. For instance, we write $x_{\agset{G}}$ to make explicit the group
$\agset{G}\subseteq Ag$ of observers of $x$.  

For each agent $a\in Ag$, the set $\var$ of variables can be partitioned into
the variables that are observable by $a$, denoted $\obs_a$, and the variable that are not observable by $a$, denoted $\nonobs_a$.
In particular, $\nonobs_{a} =\{x_\agset{G} \in \var \mid a\not\in \agset{G} \}$.
\paragraph{\bf{The base logic $\lqf$.}}
We consider a user defined base language $\lqf$, on top of which the other logics are built.
We assume that $\lqf$ is a quantifier-free first-order language with variables in $\var$.  
The Greek letter $\pi$ denotes a formula in $\lqf$.  
The following example defines $\lqf$ for integer arithmetic.
\begin{example} 
  The base language $\mathcal{L}_{\mathbb{N}}$, for integer
  arithmetic, is given by:
  \begin{align*}
    \nexpr &::  c \mid v \mid \nexpr \circ \nexpr\tag{terms} \\
    \pi &::  \nexpr=\nexpr \mid \nexpr<\nexpr \mid  \pi \wedge \pi \mid \neg \pi \tag{$\mathcal{L}_{\mathbb{N}}$ formula}
  \end{align*}
  where  $\circ::= +,-,*,/,\times, mod$; $c$ is an integer constant; and $v\in\var$.
\end{example}
We leave $\lqf$ unspecified for the rest of the paper to allow various instantiations.
\paragraph{\bf{First-order logic $\folang$}.}
We define the quantified first-order logic $\folang$
 based on $\lqf$. This logic describes ``physical'' properties of 
 a program state and serves as the target language in the translation of
 our main logic.
\begin{definition}The quantified first-order logic $\folang$ is defined by:
  \begin{align*}
    \phi &:: \pi \mid \phi \wedge \phi \mid \neg \phi \mid \forall x_\agset{G}\cdot \phi
  \end{align*}
  where $\pi$ is a quantifier-free formula in $\lqf$, and $x_\agset{G}\in \var$.
\end{definition}
Other connectives and the existential quantifier operator $\exists$, can be defined as standard.
We use Greek letters $\phi,\psi,\chi$ to denote first-order formulas in $\folang$. 
 We extend quantifiers over vectors of variables: $\forall\mathbf{x} \cdot \phi$ means $\forall x_1\cdot \forall x_2 \cdots \forall x_n\cdot \phi$. As usual,  $FV(\phi)$ denotes the set of free variables of $\phi$.

\paragraph{\bf{Epistemic logic $\epistlang$ and program-epistemic logic $\ldk$.}}
We now define two logics at once.
The first is the first-order multi-agent epistemic logic $\epistlang$ enriched with 
the public announcement operator
\footnote{The public announcement formula $[\beta]\alpha$ in the sense of \cite{plaza89,ditmarsch07} means ``after every announcement of $\beta$, $\alpha$ holds''.}.
The logic $\epistlang$ is first-order in the sense that its atomic propositions
are predicates from the base language $\lqf$.
The second is our main logic, \emph{$\ldk$}, which extends $\epistlang$
with program modalities $\square_P$.

\begin{definition}Let $\lqf$ be a base first-order language and $Ag = \{a_1, \ldots ,a_m\}$ a set of agents.
  We define the first-order  multi-agent program epistemic logic $\ldk$ with the following syntax
  \begin{align*}
    \alpha &:: \pi \mid \alpha \wedge \alpha \mid \neg \alpha \mid K_{a_i} \alpha \mid [\alpha']\alpha  \mid \forall x_\agset{G}\cdot \alpha \tag{$\epistlang$} 
  \end{align*}
  where $\pi\in\lqf$, each $K_{a_i}$ is the epistemic operator for agent $a_i$, $[\alpha']\alpha$ is a public announcement formula, $P$ is a program, $G\subseteq Ag$,
  $x_\agset{G}\in \var$, and $\forall x_\agset{G}$ is the universal quantification in the usual sense.
\end{definition}
Other connectives and the existential quantifier $\exists$,
can be defined as standard.  The program $P$ is taken from a set of
programs $\proglang$ that we define in Section~\ref{sec:programs}.  

The first-order multi-agent epistemic logic  $\epistlang$ is the fragment 
of $\ldk$ without any program operator.

\subsection{Semantics of {$\folang$} and {$\ldk$}}

\paragraph{\bf{States and the truth of {$\lqf$} formulas.}}
\label{ssub:the_state_model}

We consider a set $\domain$, used as the domain for interpreting
variables and quantifiers.  A \emph{state} $s$ of the system is a
valuation of the variables in $\var$, i.e., a function $s:\var
\rightarrow \mathsf{D}$.  We denote the universe of
all possibles states by $\mathcal{U}$.

We assume an interpretation $I$ of constants, functions, and predicates, over 
$\mathsf{D}$ to define the truth of an $\lqf$ formula at a state $s$,
denoted $s\modelsqf \pi$.

 \paragraph{\bf{Truth of an $\folang$ formula.}}
Let $s[x \mapsto c]$ denote the state $s'$ such that $s'(x) = c$ and $s'(y) = s(y)$ for all $y \in \var$ different from $x$. This lifts to a set of states, $W[x\mapsto c] = \{s[x\mapsto c] \mid s\in W\}$.   

\begin{definition}
  \label{def:truth-fol} 
The truth of $\phi\in\folang$ at a state $s$, denoted $s\modelsfo \phi$, 
is defined inductively on $\phi$ by
  \begin{alignat*}{4} 
    &s \modelsfo \pi  &&  \text{ \ iff \ } s \modelsqf \pi\\
    &s \modelsfo \phi_1 \land \phi_2 && \text{ \ iff \ }   s \modelsfo \phi_1 \text{ and } s \modelsfo \phi_2\\
    &s \modelsfo \neg \phi && \text{ \ iff \ }   s \not \modelsfo \phi\\
    &s \modelsfo \forall x_\agset{G}\cdot \phi && \text{ \ iff \  }  \text{for all } c \in \mathsf{D} , s[x_\agset{G} \mapsto c] \modelsfo \phi.
\end{alignat*}
\end{definition}
We lift the definition of $\modelsfo$ to a set $W$ of states, with $W\modelsfo \phi$ iff for all $s\in W$, $s\modelsfo \phi$. The satisfaction set $\interp{\phi}$ of a formula $\phi\in\folang$ is defined, as usual, by $\interp{\phi}=\{s\in \mathcal{U}\mid s\modelsfo \phi\}$.

{\textit{Notation.} We denote the domain of a function $f$ by $\dom(f)$,
and its post-image by $f^*$, i.e., $f^*(E) = \bigcup \{f(x)  | x\in E\}$.}

 \paragraph{\bf{Epistemic models.}}
We model agents' knowledge of the program state with a
possible worlds semantics built
on the observability of the program variables \cite{IJCAI17}. 
For that, we define, for each $a$ in $Ag$, the binary relation $\approx_a$ on $\mathcal{U}$ by:
$s\approx_a s'$ if and only if $s$ and $s'$ agrees on the part of their domains
that is observable by $a$, i.e.,
\begin{align*}
  s \approx_a s' \Leftrightarrow \textstyle
  \bigwedge_{x \in ((\dom(s)\cup\dom(s')) \cap \obs_a)} (s(x) =
  s'(x)).
\end{align*}
In particular if $s \approx_a s'$ then $\dom(s) \cap \obs_a = \dom(s') \cap \obs_a$.
Each subset $W$ of $\mathcal{U}$ defines a possible worlds model $(W, \{{\approx_{a}}_{|W}\}_{a\in Ag})$, 
  such that the states of $W$ are the possible worlds and for each $a\in Ag$
the indistinguishability relation is the restriction of $\approx_a$  on $W$.
   We shall use the set $W\subseteq \mathcal{U}$ to refer to an
epistemic model, omitting the family of equivalence relations
$\{ \approx_{a|_{W}} \}_{a\in Ag}$.

\paragraph{\bf{Truth of an $\ldk$ formula.}}
We give the semantics of an $\ldk$ formula  at a pointed model $(W,s)$, which
consist of an epistemic model $W$ and a state $s\in W$.
\begin{definition}\label{def:semantics}
Let $W$ be an epistemic model, $s\in W$ a state, $\alpha$ a formula in $\ldk$
such that $FV(\alpha) \subseteq \dom(W)$.
The truth of an epistemic
formula $\alpha$ at the pointed model $(W,s)$ is defined recursively
on the structure of $\alpha$ as follows:
  \begin{alignat*}{3} 
    (W, s) &\models \pi  &&\text{ iff \ }  s \modelsqf \pi \\
    (W, s) &\models \neg \alpha  &&\text{ iff \ }  (W, s) \not \models \alpha\\
    (W, s) &\models \alpha \land \alpha'  &&\text{ iff \ }  (W, s) \models \alpha \text{ and } (W,s) \models \alpha'\\
    (W, s) &\models K_a \alpha &&\text{ iff \  for all }s' \in W, s' \approx_{A} s \text{ implies } (W, s') \models \alpha\\
    (W, s) &\models [\beta] \alpha  &&\text{ iff \ } (W,s) \models \beta \text{ implies } (W_{|\beta},s \models \alpha)   \\
    (W,s) &\models \square_P \alpha  && \text{ iff\ } \text{for all $s'\in R_W(P,s)$, } (R^*_W(P,W),s') \models \alpha  \\
    (W,s)  &\models \forall x_\agset{G}\cdot \alpha && \text{ iff \  }  {\text{for all } c \in \mathsf{D} , \textstyle(\bigcup_{d\in \domain}
    \{s'[x_\agset{G} \mapsto d] \mid s'\in W\},s[x_\agset{G} \mapsto c]) \models \alpha}
\end{alignat*}
where $x_\agset{G}\not\in \dom(W)$,
$W_{|\beta}$ is the submodel of $W$ that consists of the states in which $\beta$ is true, i.e., $W_{|\beta}  = \{s \in \mathcal{W} \mid (W, s) \models \beta \}$~\textnormal{\cite{blackburn2006handbook}}.
\end{definition}
Our interpretation of logical connectors, epistemic formulas, and the
public announcement formulas are all standard~\cite{blackburn2006handbook,ditmarsch07}. 
In our interpretation of $\square_P\alpha$, 
the context $W$ is also updated by the relation $R_W$,
by taking the post-image of $W$ by $R_W$. The truth of $\alpha$
is interpreted at a post-state $s'$ under the new context.

For universal quantification, the epistemic context $W$ is augmented by allowing $x_\agset{G}$ to be any possible value in the domain.
When interpreting $\forall x_\agset{G} \cdot K_a \alpha'$ where $a\in \agset{G}$, we have $s\approx_a s'$ iff $s[x_\agset{G}\mapsto c] \approx_a s'[x_\agset{G}\mapsto c]$.  However, if $a\not\in \agset{G}$, then $s[x_\agset{G}\mapsto c] \approx s'[x_\agset{G}\mapsto d]$ for any $d\in \domain$ and for any $s'\approx s$. 

To interpret the formula $\square_P \alpha$, we use the function $R_W(P,\cdot):
\mathcal{U}\rightarrow\mathcal{P}(\mathcal{U})$, which models the program $P$.  
We give the function $R_W(P,\cdot)$ concretely for each
command $P$, after we define the programming language $\proglang$ in the next section.
\begin{remark}\label{R_index} 
  The index $W$ of the relational model $R_W(P,\cdot)$ of a program $P$,
  is a set of states in $\mathcal{U}$. 
  Similarly to the classical relational semantics, $R_W(P,s)$ gives the set of states resulting from executing $P$ at a state $s$.  
  However, we need the index $W$ to represent the epistemic context in which $P$ is executed.
  Before executing program $P$, an agent may not
  know that the actual initial state is $s$, it only knows about the
  initial state only as far as it can see from its observable variables.
  For that the context $W$ contains all states, which some agent may
  consider as possibly the initial states.  
\end{remark}
We define the truth of an $\ldk$ formula at an epistemic model $W$, denoted $W\models\alpha$, by $W \models \alpha \text{ iff }\text{ for any $s$ in $W$, } (W,s)\models \alpha$.

\vspace*{-0.2cm}
\section{Programming Language $\proglang$} 
\label{sec:programs}

We now give the syntax of our programs. 

We still use the notations from the previous section: \emph{$a,b,c, ...$} to denote agents,  \emph{$Ag$} to denote their whole set, \emph{$\agset{G}$} for a subset therein,  etc.

For that, we assume that a non-empty subset $\pvar$ of the set of variables $\var$ 
consists of program variables.
\begin{definition}\label{prog-lang} The programming language $\proglang$ is defined in BNF as follows:
\begin{align*}
    P ::    \varphi? \mid x_G:=e \mid \new k_G \cdot P \mid P; Q \mid P \nondet  Q
  \end{align*}
  where $x_G\in \var$, $e$ is a term over $\lqf$, $\varphi\in \epistlang$, and any variable in $P$ that is not bound by $\new\!\!$ is in $\pvar$. 
\end{definition}
The test $\varphi?$ is an assumption-like test, i.e. it blocks the program when $\varphi$ is refuted and let the program continue when $\varphi$ holds; $x_G:=e$ is a variable assignment as usual.
The program $\new k_G \cdot P$ declares a new variable $k_G$
observable by agents in $G$ before executing $P$. The operator $P;Q$ is the sequential composition of $P$ and $Q$. Lastly, $P\nondet Q$ is the nondeterministic choice between $P$ and $Q$.

Commands such as $\Skip$ and conditional tests can be defined with $\proglang$, e.g.,
\begin{alignat*}{2} 
  \If \varphi \Then P \Else Q\quad &\eqdef \quad (\varphi ?;\;P) \nondet (\neg\varphi?;\;Q).
\end{alignat*}

\paragraph{\bf{Relational semantics for $\proglang$.}}\label{sub:relational_semantics}

Now, we give the semantics of our programmes in  $\proglang$.
We refer to as classical program semantics, the modelling of a program as an input-output functionality,
without managing what agents can learn during an execution.
In classical program semantics, a program ${P}$ 
is associated with a relation $R_{{P}} = \mathcal{U}\times \mathcal{U}$, or equivalently a function $R({P},\cdot):\mathcal{U}\rightarrow \mathcal{P}(\mathcal{U})$, such that
 $R({P},\cdot)$ maps an initial state $s$ to a set of possible final states. 

As per Remark \ref{R_index}, we define the relational semantics of an epistemic program $P\in \proglang$ at a state $s$ for a given context $W$, with $s\in W$.
The context $W\subseteq\mathcal{U}$ contains states that some agents may consider as a possible alternative to $s$.
\begin{definition}[Relational semantics of $\proglang$ on states] 
  Let $W$ be a set of states. 
  The relational semantics of a program $P$ given the context $W$, is a function $R_W(P,\cdot): \mathcal{U}\rightarrow\mathcal{P}(\mathcal{U})$
  defined inductively on the structure of $P$ by
  \begin{alignat*}{2} 
      &R_W(P ; Q,s)        &&\ =\ \textstyle \bigcup_{s' \in R_W(P,s)}  \{R_{R^*_W(P,W)} (Q,s')\}  \\
      & R_W(P \nondet Q,s) && \ =\  \{s'[c_{Ag}\mapsto l] \mid s'\in R_W(P,s)\}\notag\\ 
                              & && \quad \cup  \{s'[c_{Ag}\mapsto r] \mid s'\in R_W(Q,s)\} \\
      &R_W(\new k_G\cdot P,s)        &&\ =\ R^*_W(P, \{s[k_G\mapsto d] \mid d\in\domain\})\\
      &R_W(\beta ?,s)     && \ =\  \text{if } (W,s)\models \beta \text{ then } \{s\} \text{ else } \varnothing \label{def:rel-test}\\
      &R_W(x_G:= e,s)        &&\ =\ \{s[k_G\mapsto s(x), x_G\mapsto s(e)]\}
  \end{alignat*} 
  where $k_G$ and $c_{Ag}$ are variables not in $\dom(s)$. 
\end{definition}
We model nondeterministic choice $P \nondet Q$ as a disjoint union~\cite{blackburn01}, 
which is achieved by augmenting every updated state with a new variable $c_{Ag}$, and assigning it 
 a value $l$ (for left) for every state in $R_W(P,W)$,
and a value $r$ (for right) for every state in $R_W(Q,W)$. 
The semantics of the assignment $x_G:=e$ stores the past value of $x_G$ into a new variable $k_G$,
and update the value of $x_G$ into the  $e$.
With this semantics, an agent always remembers the past values of a variable that it can observe,
i.e., it has perfect recall.
The semantics of $\new x\cdot P$ adds the new variable $x$ and allow it to take any possible value in $\domain$.

In the epistemic context, we can also view a program as transforming epistemic models,
rather than states. This view is modelled with the following alternative
relational semantics for $\proglang$.
\begin{definition}[Relational semantics of $\proglang$ on epistemic models] 
  The relational semantics on epistemic models of a program $P$ is a function 
  $F(P,\cdot) : \mathcal{P}(\mathcal{U}) \rightarrow \mathcal{P}(\mathcal{U})$ given by   
  \begin{alignat*}{2} 
      &F(x_G:= e,W)        &&\ =\ \{s[k_G\mapsto s(x_G), x_G\mapsto s(e)] \mid s\in W \}            \\
      &F(P ; Q,W)        &&\ =\   F(Q, F(P,W) )  \\
      &F(\new k_G\cdot P,W)        &&\ =\ F(P, \textstyle\bigcup_{d\in\domain} W[k_G\mapsto d])\\
      & F(P \nondet Q,W) && \ =\  \{s[c_{Ag}\mapsto l] \mid s\in F(P,W)\}\notag \\
      & &&\ \cup\ \,\{s[c_{Ag}\mapsto r] \mid s\in F(Q,W)\}  \\
      &F(\beta ?,W)     && \ =\   \{s\in W \mid (W,s) \models \beta  \} 
  \end{alignat*} 
  such that $k_G$ and $c_{Ag}$ are variables not in $\dom(s)$.
\end{definition}
The two types of relational semantics given above are related by the following.
\begin{proposition}
  For any program $P\in \proglang$ and $W\in \mathcal{P}(\mathcal{U})$, we have
  \begin{align*}
    F(P,W) = R^*_W(P,W).
  \end{align*}
\end{proposition}
\ifthenelse{\boolean{long}}{ \begin{proof}
    The proof is done by induction on the structure of $P$.
    The difficult case is that of $P;Q$. We have
    \begin{align*}
      R^*_{W}(P;Q,W) &=\textstyle  \bigcup_{s\in W}  \left\lbrace \textstyle\bigcup_{s'\in R_W\!(P,s)}  \{R_{R^*_{W}(P,W)}(Q,s') \}\right\rbrace \reason{def of $R_W(P;Q, \cdot)$ }      \\
      &=\textstyle \bigcup_{s\in W}  \left\lbrace \textstyle\bigcup_{s'\in R_W(P,s)}  \{R_{F(P,W)}(Q,s') \}\right\rbrace \reason{$F(P,W)=R^*_W(P,W)$} \\
      &=\textstyle \bigcup_{s\in W}  \left\lbrace \textstyle  R^*_{F(P,W)}(Q,R_W(P,s)) \right\rbrace \reason{by ind hypothesis on $P$}\\
      &= \textstyle  R^*_{F(P,W)}(Q,R^*_W(P,W)) \reason{$R_W^*$ is the post-image of $R_W$} \\
      &= \textstyle  R^*_{F(P,W)}(Q,F(P,W))  \reason{$F(P,W)=R^*_W(P,W)$} \\
      &= \textstyle  F(Q,F(P,W)) \reason{by induction hypothesis on $Q$.\quad \qed}\\
    \end{align*}
  \end{proof}}{  The proof can be found in the long version of our paper at~\cite{BBMR22}.}

\begin{remark} 
   We assume that every additional $c_{Ag} $, in the semantics of $P\nondet Q$, 
     is observable by all agents. The value of $c_{Ag}$ allows every agent to distinguish 
     a state resulting from $P$ from a state resulting from $Q$.
     The resulting union is a disjoint-union of multi-agent epistemic models.
     It is known that disjoint-union of models preserves the truth of epistemic formulas,
     whilst simple union of epistemic models may not~\cite{blackburn01}. 
\end{remark}

\paragraph{\bf{Weakest precondition semantics for $\proglang$.}}\label{sub:wp_semantics}

  We now give another semantics for our programs, by lifting the 
  Dijkstra's classical weakest precondition predicate transformer~\cite{dijkstra76} 
  to epistemic predicates.

\smallskip
\noindent\textit{Notation.} $\alpha[x \bk t]$ substitutes $x$ by the term $t$ in $\alpha$. 
\begin{definition}
  We define the weakest precondition of a program $P$
  as the epistemic predicate transformer $wp(P,\cdot): \epistlang  \rightarrow \epistlang $ with
  \begin{alignat*}{2} 
  & wp({P ; Q},\alpha)      && \ =\   wp(P, {wp(Q,\alpha)})  \\
  & wp({P \nondet Q},\alpha) && \ =\   wp(P,\alpha) \wedge wp(Q,\alpha)  \\
  & wp({\new k_G \cdot P,\alpha})   && \ =\  \forall k_G \cdot wp(P,\alpha) \\
  & wp({\beta ?,\alpha})   && \ =\   [\beta] \alpha                                   \\
& wp({x_G:=e,\alpha})   && \ =\  \forall k_G \cdot [k_G=e](\alpha[x_G \bk k_G])
  \end{alignat*}
  such that $FV(\alpha)\subseteq \pvar$. 
\end{definition}
The definitions of $wp$ for nondeterministic choice and sequential
composition are similar to their classical versions in the literature,
and follows the original definitions in~\cite{dijkstra76}.  A similar
definition of $wp$ for a new variable declaration is also found in
\cite{morgan94}.  However, our $wp$ semantics for assignment and for test
differs from their classical counterparts.
The classical $wp$ for assignment
(substitution), and the classical $wp$ of tests (implication) are inconsistent
in an epistemic contexts~\cite{morgan06,rajaona}.
Our $wp$ semantics for test follows from the observation that an assumption-test
for a program executed publicly corresponds to a public announcement.
Similarly, our semantics of assignment involves a public announcement of the assignment being made.

\paragraph{\bf{Equivalence between program semantics.}}\label{sub:equivalence_between_the_two_program_semantics}
 Now, we show that the two program semantics are equivalent.  For
 that, we need the following lemma, which states a validity
 correspondence after renaming a variable inside a model.  Note that
 only the name of the variable changes, the group of agents that
 observes it needs to be the same.
  \begin{lemma}
    Let $W_{x_G\bk k_G}$ rename $x_G$ into $k_G$ in the states of $W$, then
    \begin{align*}
      W \models \alpha \ \text{ iff } \ W_{x_G\bk k_G} \models \alpha[x_G\bk k_G].
    \end{align*}
  \end{lemma}
The following equivalence shows that our weakest precondition
semantics is sound w.r.t.~the program relational model.
This also serves us in proving that our $wp$-based translation of an $\ldk$
formula is sound w.r.t.~the program relational model (Section~\ref{sec:translation}).
\begin{proposition}\label{wp-vs-f} 
  For every program $P$ and every formula $\alpha\in \ldk$, 
\begin{align*}
    F(P,  W) \models \alpha \quad\text{ iff }\quad W\models wp(P,\alpha).
  \end{align*}
\end{proposition}
\ifthenelse{\boolean{long}}{\begin{proof}
Case $\beta?$  
  \begin{align*}
    &W\models wp(\beta?,\alpha)\\
    \equiv\  &W\models [\beta]\alpha \reason{the definition of $wp(\beta?,\cdot) $} \\
    \equiv\  &\forall s\in W, (W,s) \models [\beta]\alpha \reason{by the definition of $\models$ on a model} \\  
    \equiv\  &\forall s\in W, \text{ if } (W,s)\models \beta \text{ then } (W_{|\beta},s)\models \alpha \reason{$\models$ for public announcement}  \\  
    \equiv\  &\forall s\in W, \text{ if } (W,s)\models \beta \text{ then } (\{s'\in W| (W,s')\models \beta\},s)\models \alpha \reason{def of $W_{|\beta}$}  \\  
    \equiv\  &\forall s\in W, \text{ if } s\in F(\beta?,W) \text{ then }  (F(\beta?),s)\models \alpha \reason{by definition of $F(\beta?,\cdot) $}  \\
    \equiv\  & F(\beta?,W)\models \alpha  \reason{by the definition of $\models$ on a model}
  \end{align*}
  Case $P\nondet Q$
  \begin{align*}
    & F(P\nondet Q,W) \models \alpha \\
    \equiv\  &  \{s[c_{Ag}\mapsto l]| s\in F(P,W)\} \cup \{s[c_{Ag}\mapsto l]| s\in F(Q,W)\}
    \models\alpha \reason{the definition of $F(P\nondet Q,\cdot)$ }  \\
    \equiv\  & \{s[c_{Ag}\mapsto l]| s\in F(P,W)\}\models\alpha \text{ and } \{s[c_{Ag}\mapsto l]| s\in F(Q,W)\}\models \alpha \reason{by Prop 2.3 in~\cite{blackburn01}, this is a disjoint union since $c_{Ag}$ observable by all}
 \\
\equiv\  & F(P,W)\models \alpha \text{ and }F(Q,W)\models \alpha \reason{ $c_{Ag}$ is not in $\alpha$}   \\
    \equiv\  & W\models wp(P,\alpha) \text{ and }W \models wp(Q,\alpha) \reason{by induction hypothesis on $P$ and $Q$} 
  \end{align*}
\raisebox{.8em}{Case $P; Q$}      \begin{minipage}[c]{.86\textwidth} 
      \begin{align*}
        F(P; Q,W) \models \alpha \equiv\ &  F(Q,F(P,W)) \models \alpha \reason{definition of $F$ for $P;Q$}\\
        \equiv\  & F(P,W) \models wp(Q,\alpha)\reason{induction hypothesis on $Q$} \\
        \equiv\  & W \models wp(P,wp(Q,\alpha))\reason{induction hypothesis on $P$}  
\end{align*}
    \end{minipage} 
Case $\new k \cdot P$
  \begin{align*}
    &\ W \models wp(\new k\cdot P, \alpha)  \\
 \equiv\  & \text{ for any $s\in W$, }  (W,s)\models \forall k\cdot wp(P,\alpha) \reason{the definition of $wp$ for $\new k$} \\
 \equiv\  & \text{ for any $s\in W$ and any $c\in D$, } \textstyle (\bigcup_{d\in \domain} W[k \mapsto d],s[k\mapsto c] )\models wp(P,\alpha) \reason{by definition of $\models$ for $\forall k$} \\
 \equiv\  & \text{ for any $s'\in\textstyle \bigcup_{d\in \domain} W[k \mapsto d]$, } \textstyle (\bigcup_{d\in \domain} W[k \mapsto d],s' )\models wp(P,\alpha) \\
 \equiv\  & \textstyle\bigcup_{d\in D} W[k\mapsto d] \models wp(P,\alpha) \reason{by lifting $\models$ to the entire model }  \\
 \equiv\  & F(P,\textstyle\bigcup_{d\in D} W[k\mapsto d]) \models \alpha \reason{by induction hypothesis on $P$}\\
 \equiv\  & F(\new k \cdot P, W ) \models \alpha \reason{the definition of $F(\new k,\cdot)$. } 
  \end{align*}
Case $x_G:=e$. 
  To understand the proof, observe that the action of $F(x_G:=e,\cdot)$ on $W$, 
  is equivalent to renaming the old $x_G$ into $k_G$,
  then making a new variable $x_G$ that takes the value $e$.
  We also need to derive the following equality
  \begin{align*}
     &F(x_G:=e, W)\\
     =\ & \{s[k_G\mapsto s(x_G), x_G\mapsto s(e)] | s\in W \} \reason{by definition of $F(x_G:=e,\cdot)$} \\
     =\ & \{s[x_G\mapsto s(e_{x_G\bk k_G})] | s\in W_{x_G\bk k_G} \} \reason{by definition of $W_{x_G\bk k_G}$} \\
     =\ & \textstyle (\bigcup_{d\in \domain} W_{x_G\bk k_G}[x_G\mapsto d])_{|d=s(e_{x_G\bk k_G})} \reason{because $x_G$ is not in $\dom(W_{x_G\bk k_G})$}  \\
     =\ & F((x_G=e_{x_G\bk k_G})?, \textstyle \bigcup_{d\in \domain} W_{x_G\bk k_G}[x_G\mapsto d]) \reason{by definition of $F$ for tests}   \\
     =\ & F(\new x_G \cdot (x_G=e_{x_G\bk k_G})?, W_{x_G\bk k_G})  \reason{by definition of $F$ for $\new x_G$.}
  \end{align*}
  where $W_{x_G\bk k_G}$ renames $x_G$ into $k_G$ in the states of $W$. Now, 
  \begin{align*}
    & F(x_G:=e, W)\models \alpha\\
     \equiv\ &F(\new x_G \cdot (x_G=e_{x_G\bk k_G})?, W_{x_G\bk k_G}) \models \alpha \reason{from the previous equality}  \\
    \equiv\ &F(\new k_G \cdot (k_G=e)?, W) \models \alpha_{x_G\bk k_G}\reason{after swapping $x_G$ and $k_G$} \\ 
    \equiv\ &W \models wp(\new k_G\cdot (k_G=e)?,\alpha_{x_G\bk k_G} )\reason{by induction hypothesis on $\new k_G$} \\
    \equiv\ &W \models \forall k_G [k_G=e] \alpha_{x_G\bk k_G} \reason{by the definition of $wp$ for assignment.\quad\qed} 
  \end{align*}
\end{proof}}{ The proof can be found in the long version of our paper at~\cite{BBMR22}.}

\begin{remark}
  By modelling nondeterministic choice as a disjoint union,
  the epistemic formulas that are $F(P,W)$ and true at $F(Q,W)$, remains true at $F(P\nondet Q, W)$.
  This allows us to have a conjunctive weakest precondition epistemic predicate transformer, 
  i.e. $wp(P\nondet Q,\alpha) =  wp(P,\alpha) \wedge wp(Q,\alpha)$.
\end{remark}

In our logics, we included the public announcement operator to
simplify some aspects of the presentation, although the public
announcement formula $[\beta]\alpha$ is equivalent to the dynamic
formula $\square_{\beta?}\alpha$, as per the following proposition.
\begin{proposition}\label{eq:box-pa} Let $\beta\in \epistlang$, then
  $(W, s) \models \square_{\beta?} \alpha  \text{\ \ iff\ \ }
  (W,s) \models [\beta] \alpha$.
\end{proposition}
\ifthenelse{\boolean{long}}{ \begin{proof}~
    For the case $(W,s)\not\models \beta$, the equivalence follows from the equivalence of $(W,s)\not\models \beta$ and $ R_W(\beta?,s) = \varnothing$. 
    Now, assume $(W,s)\models \beta$, then $R_W(\beta?,s)=\{s\}$. Then we have
        $(W,s) \models \square_{\beta?} \alpha 
    \text{ iff }   (R^*_W(\beta?,W),s) \models \alpha$.
The rest follows from $R^*_W(\beta?,W) = F(\beta?,W) =  W_{|\beta}$.\qed
\end{proof}}{The proof can be found in the long version of our paper at~\cite{BBMR22}.}

\vspace*{-0.3cm}
\section{Translating $\ldk$ to $\folang$} 
\label{sec:translation}

Our model checking approach relies on the truth-preserving translation between
$\ldk$ formulas and first-order formulas.
We use the following translation function.
\begin{definition}[Translation of $\ldk$ into $\folang$]\label{translation} 
  Let 
  $\pi\in \lqf$ and $\alpha \in \ldk$.  We defined the
translation $\tau: \folang \times \ldk \to \folang$  as follows:
  \begin{align*}
    \tau(\phi,\pi)  &\ =\   \pi \\
    \tau(\phi,\neg \alpha)  & \ =\   \neg \tau(\phi,\alpha) \\
    \tau(\phi,\alpha_1\circ\alpha_2) & \ =\   \tau(\phi,\alpha_1)\circ \tau(\phi,\alpha_2) \\
    \tau(\phi, K_a \alpha) & \ =\   \forall \mathbf{n} \cdot (\phi \rightarrow \tau(\phi,\alpha)) \\  
    \tau(\phi, [\beta] \alpha) & \ =\    \tau(\phi ,\beta) \rightarrow \tau(\phi\wedge\tau(\phi,\beta) ,\alpha  )\\
    \tau(\phi, \square_P \alpha) & \ =\    \tau(\phi,wp(P,\alpha)) \\
    \tau(\phi, \forall x_G\cdot \alpha) & \ =\   \forall x_G\cdot \tau(\phi,\alpha) 
\end{align*}
  where $\circ \in \{\land, \lor \}$ and $\mathbf{n} = \nonobs_a\cap(FV(\alpha)\cup FV(\phi))$.
\end{definition}

We use the above translation to express the equivalence between
the satisfaction of a $\epistlang$-formula and that of its first-order translation.
\begin{proposition}\label{translation-pointwise} 
    For every $\phi$ in $\mathcal{L}_{FO}$, $s$ in $\interp{\phi} $, $\alpha$ in $\epistlang$ such that
    $FV(\phi)\cup FV(\alpha)\subseteq \pvar$, we have that
    \begin{align*}
      (\interp{\phi},s) \models \alpha \text{ iff } s \modelsfo \tau(\phi,\alpha).   
    \end{align*}
\end{proposition}
\ifthenelse{\boolean{long}}{\begin{proof}~
    The proof for the base epistemic logic without public announcement $\mathcal{L}_K$ ($\pi, \neg, \wedge, K_a$) is found in~\cite{IJCAI17}.

    \noindent Case of public announcement $[\beta]\alpha$
    \begin{align*}
      &(\interp{\phi},s) \models [\beta]\alpha \\
      \equiv \ &\text{if } (\interp{\phi},s) \models \beta \text{ then } (\interp{\phi}_{|\beta} ,s) \models \alpha \reason{truth of $[\beta]\alpha$ }   \\
      \equiv \ &\text{if } s\modelsfo \tau(\phi,\beta) \text{ then } (\interp{\phi}_{|\beta} ,s) \models \alpha \reason{induction hypothesis on $\beta$}  \\
      \equiv \   &\text{if } s\modelsfo \tau(\phi,\beta) \text{ then }  (\{s'\in \mathcal{U}| s'\modelsfo \phi \text{ and }  (\interp{\phi},s')\models \beta \},s) \models \alpha \reason{by definition of $\interp{\cdot}$ and definition of ${_{|\beta}}$ }  \\
                      \equiv\  &\text{if } s\modelsfo \tau(\phi,\beta) \text{ then }  (\{s'\in \mathcal{U}| s'\modelsfo \phi \text{ and } s'\modelsfo \tau(\phi,\beta) \},s) \models \alpha \reason{induction hypothesis on $\beta$}\\
                      \equiv\  &\text{if } s\modelsfo \tau(\phi,\beta) \text{ then }  (\{s'\in \mathcal{U}| s'\modelsfo \phi \wedge\tau(\phi,\beta) \},s) \models \alpha \reason{truth of $\wedge$} \\
                      \equiv\  &\text{if } s\modelsfo \tau(\phi,\beta) \text{ then }  (\interp{\phi\wedge\tau(\phi,\beta)} ,s) \models \alpha \reason{def of $\interp{\cdot}$} \\
                      \equiv\  &\text{if } s\modelsfo \tau(\phi,\beta) \text{ then } s\modelsfo\tau(\phi\wedge \tau(\phi,\beta),\alpha)\reason{induction hypothesis}  \\  
                      \equiv\  &\text{if } s\modelsfo \tau(\phi,\beta) \rightarrow \tau(\phi\wedge \tau(\phi,\beta),\alpha)  \reason{truth of $\rightarrow$.\quad\qed}
    \end{align*}

\end{proof}
}{
  The proof can be found in the long version of our paper~\cite{BBMR22}.
}

Now, we can state our main theorem relating the validity of an $\ldk$ formula,
and that of its first-order translation.
\begin{theorem}[Main result] \label{main} 
    Let $\phi\in \mathcal{L}_{FO}$, and $\alpha \in \ldk$, such that
    $FV(\phi)\cup FV(\alpha)\subseteq \pvar$, then
    \begin{align*}
      \interp{\phi} \models \alpha \text{ iff } \interp{\phi} \modelsfo \tau(\phi,\alpha).
    \end{align*}
  \end{theorem}
\ifthenelse{\boolean{long}}{
  \begin{proof} The proof is done by induction on $\alpha$. We start with $\alpha\in\epistlang$.
    \begin{align*}
     \interp{\phi} \models \alpha \equiv\ &\text{ for all $s$ in $\interp{\phi}$, }  (\interp{\phi},s) \models \alpha \reason{by definition of $\models$ for a model} \\
      \equiv\ &\text{ for all $s$ in $\interp{\phi}$, }  s \modelsfo \tau(\phi,\alpha)\reason{by Proposition \ref{translation-pointwise}} \\
      \equiv\ &\interp{\phi} \modelsfo \tau(\phi,\alpha) \reason{by definition of $\modelsfo$ for a set of states}  
    \end{align*}

    Now, we prove for the case of the program operator $\square_P\alpha$. 
    Without loss of generality, we can assume that $\alpha$ is program-operator-free, i.e., $\alpha\in \epistlang$.
    Indeed, one can show that $\square_P(\square_Q \alpha')$ is equivalent to $\square_{P;Q}\alpha'$. 
We have
    \begin{align*}
      &\interp{\phi} \models \square_P\alpha \\ 
      \equiv\ &\text{ iff for all $s$ in $\interp{\phi}$, } (\interp{\phi},s) \models \square_P\alpha \reason{by definition of $\models$ for a model}\\ 
      \equiv\ &\text{ iff for all $s$ in $\interp{\phi}$, for all $s'$ in $R_{\interp{\phi}} (P,s)$, } (F(P,\interp{\phi}),s') \models \alpha \reason{by definition of $\models$ for $\square_P$}\\
      \equiv\ &\text{ iff for all $s'$ in $R^*_{\interp{\phi}}  (P,\interp{\phi})$, } (F(P,\interp{\phi}),s') \models \alpha \reason{post-image} \\
      \equiv\ &\text{ iff for all $s'$ in $F  (P,\interp{\phi})$, } (F(P,\interp{\phi}),s') \models \alpha\reason{$F(P,W) = R^*_W(P,W)$ }  \\
      \equiv\ &F(P,\interp{\phi}) \models \alpha \reason{by definition of $\models$ for a model}\\
      \equiv\ &\interp{\phi} \models wp(P,\alpha)\reason{by Proposition \ref{wp-vs-f}} \\
      \equiv\ &\interp{\phi}\modelsfo \tau(wp(P,\alpha)) \reason{since $wp(P,\alpha)\in \epistlang$, the previous case applies. \qed}  
    \end{align*}
  \end{proof}
}{
  The proof can be found in the long version of our paper~\cite{BBMR22}.
}

\vspace*{-0.3cm}
\section{Implementation} 
\label{sec:implementation} 
Our automated verification framework supports proving/falsifying a logical consequence $\phi\models \alpha$ 
for $\alpha$ in $\ldk$ and $\phi$ in $\folang$.
By Theorem~\ref{main}, the problem becomes the unsatisfiability/satisfiability of first-order formula $\phi \wedge \neg\tau(\phi, \alpha)$, which is eventually fed to an SMT solver.

In some cases, notably our second case study, the Cheryl's Birthday puzzle,
 computing the translation $\tau(\phi, \alpha)$ by hand is tedious and error-prone. 
For such cases, we implemented a $\ldk$-to-$\folang$ translator to automate the translation.

\subsection{Mechanisation of Our $\ldk$-to-FO Translation}\label{ssub:mechanised_translation}
Our translator implements Definition~\ref{translation} of our translation $\tau$.
It is implemented in \texttt{Haskell}, and it is generic,
i.e., works for any given example\footnote{Inputs are \texttt{Haskell}
files.}.
The resulting  first-order formula is exported as a string
 parsable by an external SMT solver API
(e.g., \texttt{Z3py} and \texttt{CVC5.pythonic}  which we use).

Our \texttt{Haskell} translator and the implementation of our case studies
are available at \url{https://github.com/UoS-SCCS/program-epistemic-logic-2-smt}.

\subsection{Case Study 1: Dining Cryptographers' Protocol~\cite{Chaum1988}. }\label{sub:dining_cryptographer_protocol}

\paragraph{\bf{Problem Description.}} This system is described by $n$ cryptographers dining round a table. 
One cryptographer may have paid for the dinner, or their employer may
have done so. They execute a protocol to reveal whether one of the
cryptographers paid, but without revealing which one. Each pair of
cryptographers sitting next to each other have an unbiased coin, which
can be observed only by that pair. Each pair tosses its coin. Each
cryptographer announces the result of XORing three booleans: the two
coins they see and the fact of them having paid for the dinner. The
XOR of all announcements is provably equal to the disjunction of
whether any agent paid.

\paragraph{\bf{Encoding in $\ldk$ \& Mechanisation.} }We consider the domain $\mathbb{B}=\{T,F\}$ and the program variables
\mbox{$\pvar = \{x_{Ag}\} \cup \{ p_i, \allowbreak c_{\{i,i+1\}} \mid 0\le i< n\}$} where
$x$ is the XOR of announcements; $p_i$ encodes whether agent~$i$ has paid;
and, $c_{\{i,i+1\}}$ encodes the coin shared between agents~$i$ and~$i+1$.
The observable variables for agent \mbox{$i\in Ag$}  are
\mbox{$\obs_{i}  = \{x_{Ag}, p_i, c_{\{i-1,i\}}, c_{\{i,i+1\}} \}$}
\footnote{
  When we write $\{i,i+1\}$ and $\{i-1,i\}$, we mean $\{i,i+1 \bmod
n\}$ and $\{i-1 \bmod n,i \}$.  } ,
and \mbox{$\nonobs_i=\pvar\setminus \obs_{i}$}.

We denote $\phi$ the constraint that at most one agent have paid,
and $e$ the XOR of all announcements, i.e.
\begin{equation*}
  \textstyle \phi = \bigwedge_{i=0}^{n-1} \left(p_i \Rightarrow \bigwedge_{j=0,j\neq i}^{n-1} \neg p_j\right) \qquad e = {\textstyle\bigoplus_{i=0}^{n-1} p_i\oplus c_{\{i-1,i\}} \oplus c_{\{i,i+1\}}}.
\end{equation*}
The Dining Cryptographers' protocol is modelled by the program
  $P \ =\  \textstyle {x_{Ag}}:= e$.

\paragraph{\bf{Experiments \& Results.} }
We report on checking the validity for: 
\begin{alignat*}{4} 
       \textstyle \beta_1&=\textstyle\square_P \left((\neg p_0) \Rightarrow \left(K_0\left(\bigwedge_{i=1}^{n-1}\neg
        p_i\right)\vee \bigwedge_{i=1}^{n-1} \neg K_0 p_i \right)\right) \quad \beta_3 = \textstyle  \square_P (K_0 p_1) \\
       \textstyle \beta_2 & = \textstyle \square_P\left( K_0\left(x \Leftrightarrow \bigvee_{i=0}^{n-1}p_i\right)\right) \qquad\qquad\   \gamma =\textstyle K_0 \left(\square_P \left(x \Leftrightarrow \bigvee_{i=0}^{n-1}p_i\right)\right). 
      \end{alignat*}
The formula $\beta_1$ states that after the program execution, if
cryptographer~$0$ has not paid then she knows that no cryptographer
paid, or (in case a cryptographer paid) she does not know which
one. 
The formula $\beta_2$ reads that after the program execution, cryptographer~$0$ knows that $x_{Ag}$ is true iff one of the cryptographers paid.
The formula $\beta_3$ reads that after the program execution, cryptographer~$0$ knows that cryptographer~$1$ has paid, which is expected to be false.
Formula $\gamma$ states cryptographer~$0$ knows that, at the end of
 the program execution, $x_{Ag}$ is true iff one of the cryptographers
 paid.

Formulas $\beta_1,\beta_2,$ and $\beta_3$ were checked in~\cite{IJCAI17} as well.
Importantly, formula $\gamma$
cannot be expressed or checked by the framework in~\cite{IJCAI17}.
We compare the performance of our translation on this case-study with that of~\cite{IJCAI17}. 
To fairly compare, we reimplemented faithfully  the SP-based translation 
in the same environment as ours.
We tested our translation (denoted $\tau_{\wp}$) and the reimplementation
of the  translation in~\cite{IJCAI17} (denoted $\tau_{\SP}$) on the same
machine.

Note that the performance we got for $\tau_\SP$ differs
from what is reported in~\cite{IJCAI17}. 
This is especially the case for the most complicated formula $\beta_1$.
This may be due to the machine
specifications, or because we used binary versions of $\texttt{Z3}$ and $\texttt{CVC5}$,
rather than building them from source, like in~\cite{IJCAI17}.

The results of the experiments, using the \texttt{Z3} solver, are shown in Table~\ref{table}. 
\texttt{CVC5} was less performant than \texttt{Z3} for this example, as shown (only) for $\beta_2$.
Generally, the difference in performance between the two translations were small.
The $\SP$-based translation slightly outperforms our translation for $\beta_2$ and $\beta_3$,
but only for some cases.
Our translation outperforms the $\SP$-based translation for $\beta_1$ in these experiments.
Again, we note that the performance of the $\SP$-based translation reported here
is different from the performance reported in \cite{IJCAI17}.
Experiments that took more than 600 seconds were timed out

\newcommand{\mcol}[3]{\multicolumn{#1}{#2}{#3}}   
\newcommand{\spZ}{{$\tau_{\textit{SP}}$+\texttt{Z3} }}  
\newcommand{\spZnil}{{$\tau_{\textit{SP}}$ }}  
\newcommand{\wpZ}{{$\tau_{\textit{wp}}$+\texttt{Z3}}}  
\newcommand{\wpC}{{$\tau_{\textit{wp}}$+\texttt{CVC5}}}  

\begin{table}[htpb] 
 \begin{small}
 \begin{center}
 \begin{tabular}{c@{\ \ }c@{\ }c@{\ \ }c@{\ }c@{\ \ }c@{\ }c@{\ \ }c@{\ }c@{\ \ }c@{\ }c@{\ }} 
 \toprule
& \mcol{2}{c}{Formula $\beta_{1}$} & \mcol{3}{c}{Formula $\beta_2$} & \mcol{2}{c}{Formula $\beta_3$} & \mcol{2}{c}{Formula $\gamma$}        \\ \cmidrule(lr){2-3} \cmidrule(lr){4-6} \cmidrule(lr){7-8} \cmidrule(lr){9-10}
 n   & \wpZ                             & \spZ                     & \wpC     & \wpZ                           & \spZ   & \wpZ   & \spZ   & \wpZ   & \spZ\\ \cmidrule(l){2-2}\cmidrule(l){3-3}\cmidrule(l){4-4}\cmidrule(l){5-5}\cmidrule(lr){6-6}\cmidrule(lr){7-7}\cmidrule(lr){8-8}\cmidrule(lr){9-9}\cmidrule(lr){10-10}
 10  & 0.05 s                           & 4.86 s                   & 0.01 s   & 0.01 s                         & 0.01 s & 0.01 s & 0.01 s & 0.01 s & {N/A}         \\
 50  & 31   s                           & t.o.                     & 0.41 s   & 0.05 s                         & 0.06 s & 0.03 s & 0.02 s & 0.03 s & {N/A} \\
 100 & t.o.                             & t.o.                     & 3.59 s   & 0.15 s                         & 0.16 s & 0.07 s & 0.06 s & 0.07 s & {N/A}\\
 200 & t.o.                             & t.o.                     & 41.90 s   & 1.27 s                         & 0.71 s & 0.30 s & 0.20 s & 0.30 s & {N/A}
 \end{tabular}
 \end{center}
 \end{small}
 \caption{Performance our $\wp$-based translation vs. our reimplementation of the~\cite{IJCAI17} $\SP$-based translation for the Dining Cryptographers.
 Formula $\gamma$ is not supported by the $\SP$-based translation in ~\cite{IJCAI17}.} 
  \label{table} 
\end{table}
 
\subsection{Case Study 2: Cheryl's Birthday Puzzle~\cite{van2017cheryl}.}\label{sub:cheryl_s_birthday}
This case study involves the nesting of knowledge operators $K$
of different agents. 

\paragraph{\bf{Problem Description.}}
Albert and Bernard just became friends with Cheryl, and they want to
know when her birthday is.  Cheryl gives them a list of 10 possible
dates: May 15, May 16, May 19, {June} 17, {June} 18, July 14, July 16,
{August} 14, {August} 15, {August} 17.  Then, Cheryl whispers in
Albert’s ear the month and only the month of her birthday.  To
Bernard, she whispers the day only. ``Can you figure it
out now?'', she asks Albert.  The next dialogue follows:

\begin{raggedright}
  \quad - Albert: I don’t know when it is, but I know Bernard doesn’t know either.\\
  \quad - Bernard: I didn’t know originally, but now I do.\\
  \quad - Albert: Well, now I know too!\\
\end{raggedright}
\noindent When is Cheryl’s birthday?

\paragraph{\bf{Encoding and Mechanisation.}} 
To solve this puzzle, we consider two agents $a$ (Albert) and $b$ (Bernard)  and two integer program variables
$\pvar = \{m_a, d_b\}$.
Then, we constrain the initial states to satisfy the conjunction of all possible dates announced by Cheryl, i.e.,  the formula $\phi$  below:  \begin{align*}
    \phi(m_a,d_b) =&\  (m_a = 5 \wedge d_b = 15) 
    \vee(m_a = 5 \wedge d_b = 16) \vee \ \cdots
\end{align*}
The puzzle is modelled via public announcements, with the added assumption that participants  tell the truth. 
However, modelling a satisfiability problem with the public announcement operator $[\beta]\alpha$ would return states where $\beta$ cannot be truthfully announced.
Indeed, if $\beta$ is false at a $s$, (i.e $(\phi,s)\models \neg\beta$, 
then the announcement $[\beta]\alpha$ is true. 
For that, we use the dual 
of the public announcement operator
denoted  $\langle\cdot\rangle$ 
\footnote{The formula $\langle\beta\rangle\alpha$ reads ``after some announcement of $\beta$, $\alpha$ is the case'', i.e., $\beta$ can be truthfully announced and its announcement makes $\alpha$ true. 
  Formally, $  (W, s) \models \langle\beta\rangle \alpha  \text{ iff } (W,s) \models \beta \text{ and } (W _{|\beta},s) \models \alpha$.}.
We use the translation to first-order formula:
\begin{align*}
  \tau(\phi, \langle\beta\rangle \alpha) & \ =\    \tau(\phi ,\beta) \wedge \tau(\phi\wedge\tau(\phi,\beta) ,\alpha  ).
\end{align*}
In both its definition and our translation to first-order,  $\langle\cdot\rangle$ uses a conjunction
where $[\cdot]$ uses an implication.

We denote the statement ``agent $a$ knows the value of $x$'' by the formula 
$\kv_a x$ which is common in the literature. We define it with our logic $\ldk$
making use of existential quantification:
  $\kv_a x \ =\ \exists v_a \cdot K_a (v_a = x)$. 

Now, to model the communication between Albert and Bernard,
let $\alpha_a$ be Albert's first announcement, i.e., $\alpha_a = \neg \kv_a (d_b) \wedge K_a(\neg \kv_b (m_a))$.
Then, the succession of announcements by the two participants correspond to the formula 
\begin{align*}
  \alpha = \langle{(\neg \kv_b (m_a)\wedge {\langle \alpha_a \rangle}  \kv_b (m_a) )?\rangle} \kv_a d_b.
\end{align*}

Cheryl's birthday is the state $s$ that satisfies $(\phi,s) \models \alpha$.

\subsubsection{Experiments \& Results.}

We computed $\tau(\phi,\alpha)$ with our translator in 0.10 seconds.
The SMT solvers \texttt{Z3} and \texttt{CVC5} returned the solution to the puzzle when fed with $\tau(\phi,\alpha)$.
\texttt{CVC5} solved it, in 0.60 seconds, 
which is twice better than \texttt{Z3} (1.28 seconds). 

All the experiments were run on a 
6-core 2.6 GHz Intel Core i7 MacBook Pro with 16 GB of RAM running OS X 11.6.
For \texttt{Haskell}, we used GHC 8.8.4. 
The SMT solvers were \texttt{Z3} version 4.8.17 and \texttt{CVC5} version 1.0.0.

\vspace*{-0.3cm}
\section{Related Work}

\paragraph{\bf{SMT-Based Verification of Epistemic Properties of Programs.}} We start with the work of Gorogiannis~\etal~\cite{IJCAI17} which is the closest to ours. We already compared with this in the introduction, for instance explaining therein exactly how our logic is much more expressive than theirs. Now, we cover other points.

\textit{Program Models.} The program models in \etal\cite{IJCAI17} follow a classical program semantics (e.g., modelling nondeterministic choice as union, overwriting a variable in reassignment). This has been shown~\cite{morgan06,rajaona} to correspond to systems where agents have no memory, and cannot see how nondeterministic choices are resolved. 
Our program models assume perfect recall, and  that agents can see how nondeterministic choices are resolved.

\textit{Program Expressiveness.} Gorogiannis \etal\cite{IJCAI17} have results of approximations for programs with loops, although there were no use cases of that. Here we focused on a loop-free programming language, but we believe our approach can be extended similarly. The main advantage of our programs is the support for tests on knowledge which allows us to model public communication of knowledge.

\textit{Mechanisation \& Efficiency.} We implemented the translation which include an automated computation of weakest preconditions (and strongest postconditions as well). The implementation in~\cite{IJCAI17} requires the strongest postcondition be computed manually.  Like~\cite{IJCAI17}, we test for the satisfiability of the resulting first-order formula with \texttt{Z3}. The performance is generally similar, although sometimes it depends on the form of the formulas (see Table \ref{table}).  

\paragraph{\bf{Refinement-based Verification of Epistemic Properties of Programs.}}
Verifying epistemic properties of programs via refinement  was done in~\cite{morgan06,mciver09b,rajaona}. Instead of using a dynamic logic, they reason about epistemic properties of programs with an ignorance-preserving refinement. 
Like here, their notion of knowledge is based on observability of arbitrary domain program variables. 
The work in~\cite{rajaona} also consider a multi-agent logics and nested $K$ operators and their program also allows for knowledge tests. Finally, our model for epistemic programs can be seen as inspired by~\cite{rajaona}.  That said, all these work have no relation with first-order satisfaction nor translations of validity of programme-epistemic logics to that, nor their implementation.

\paragraph{\bf{Dynamic Epistemic Logics}}    \textit{Dynamic epistemic logic} (DEL, \cite{plaza89,baltag99,ditmarsch07}) is a family of logics that extend epistemic logic with dynamic operators.

\textit{Logics' Expressivity.}
On the one hand, DEL logics are mostly propositional, and 
their extensions with assignment only considered propositional assignment (e.g.,~\cite{ditmarsch05}); contrarily, we support assignment on variables on arbitrary domains. Also, we have a denotational semantics of programmes (via weakest preconditions), whereas DEL operates on more abstract semantics.
On the other hand, action models in DEL can describe complex  private communications
that cannot be encoded with our current programming language.

\textit{Verification.}  Current DEL model checkers include \texttt{DEMO}~\cite{van2007demo} and \texttt{SMCDEL}~\cite{van2015symbolic}.
We are not aware of the verification of DEL fragments being reduced to satisfiability problems. In this space, an online report~\cite{DELreport} discusses --at some high level-- the translation  \texttt{SMCDEL} knowledge structures into QBF and the use of \texttt{YICES}. 

A line of research in DEL, the so called \textit{semi-public environments}, also builds agents' indistinguishability relations from the observability  of propositional variables~\cite{wooldridge01b,charrier2016building,semi-public}.
The work of Grossi~\cite{grossi2017non-determinism} explores the interaction between knowledge dynamics and non-deterministic choice/sequential composition. 
They note that PDLs assumes memory-less agents and totally private nondeterministic choice, whilst DELs' epistemic actions assume agents with perfect recall and publicly made nondeterministic choice. 
This is the same duality that we observed earlier between the program epistemic logic in~\cite{IJCAI17} and ours. 

\paragraph{\bf{Other Works.}} Gorogiannis \etal~\cite{IJCAI17}  discussed more tenuously related work, such as on general verification of temporal-epistemic properties of systems which are not programs in  tools like \texttt{MCMAS}~\cite{mcmas}, \texttt{MCK}~\cite{mck}, \texttt{VERICS}~\cite{verics}, or one line of epistemic verification of models specifically of \texttt{JAVA} programs~\cite{encover}. \cite{IJCAI17} also discussed some incomplete method of SMT-based  epistemic model checking~\cite{Cimatti:2016}, or even bounded model checking techniques, e.g.,
\cite{kacprzak2006comparing}. All of those are loosely related to us too, but there is little reason to reiterate.

\vspace*{-0.3cm}
\section{Conclusions}

We advanced a multi-agent epistemic logics for programs $\ldk$, 
in which each agent has visibility over some program variables but not others.
This logic allows to reason on agents' knowledge of a program after its run,
as well as before its execution.  
Assuming agents' perfect recall, we provided 
a weakest-precondition epistemic predicate transformer semantics
that is sound w.r.t to its relational counterpart.
Leveraging the natural correspondence between the weakest precondition $wp(P,\alpha)$
and the dynamic formula $\square_P\alpha$,
we were able to give a sound reduction of the validity of $\ldk$ formulas to first-order satisfaction. 
We provided an implementation of this  $\ldk$ verification methods 
and report on the verification of series of benchmarks from the literature.
The multi-agent nature of the logic, the expressiveness of it w.r.t. knowledge evaluation 
before and after program execution, as well as a complete verification method 
for this are all novelties in the field.

In future work, we will look at a meet-in-the-middle between the memoryless semantics in~\cite{IJCAI17} 
and the   memoryful semantics here, and methods of verifying logics like $\ldk$ 
but with such less ``absolutist" semantics.

\end{document}